# An Accurate Measurement System Comprising of Wireless Thermometers for Neonate Body Temperature Monitoring

Javadpour Abdollah[1], Rahimi A.[2]*, Javadpour Amir[3], Vazini H.[4], Farhadi Z.[5], Baghdadpour Gh. H.[6], Mardani Korani Z.[7]

[1]Health Research Center, Baqiyatallah University of medical sciences, Tehran, Iran
[2]Health Research Center, Baqiyatallah University of medical sciences, Tehran, Iran
[3]Faculty of New Sciences & Technologies, University of Tehran, Tehran, Iran
[4]Department of Nursing, Hamedan Branch, Islamic Azad University, Hamadan, Iran
[5]Nursing and Midwifery educational office, Babol University of Medical Sciences, Babol, Iran
[6]Department of Computer Electronic ITMC, Applied Science Shiraz University, Shiraz, Iran
[7]Department of Computer Engineering, Khorasgan (Isfahan) Branch, Islamic Azad University, Isfahan, Iran

*Corresponding author:
A. Rahimi
Health Research Center, Baqiyatallah University of medical sciences, Tehran, Iran
E-mail: Fazel123@yahoo.com

## ABSTRACT

This paper addresses the design and implementation of a real-time monitoring system consisting of wireless thermometers (wireless temperature sensor node) for continuous recording of neonate body temperature in intensive care unit (ICU). Each wireless thermometer incorporates an accurate semiconductor temperature sensor, a transceiver operating on ISM frequency band (i.e., 915 MHz) as well as a microcontroller which is used to control the thermometers functionalities including wireless medium access (e.g., free space or body channel), transmission and reception. A voltage regulator and a low-pass filter were also used for removing spurious signals and environmental noise from the thermometer feed line. In order to distinguish the reading of each thermometer from measurements performed by other thermometers, the I-wire protocol was used. This protocol can securely tag the temperature value by incorporating a unique ID (provided by the manufacturer) into the packet sent wirelessly. An array of two thermometers was implemented and successfully tested in different scenarios, namely free-space, water (immersed thermometers) and on a volunteer's wrist. Moreover, an in-house developed computer software was used in order to visualize the readings in addition to alerting rapid increase and high body temperature. The software also compares the measurement results with actual values. The agreement between the experimental data and real temperature values is reasonably acceptable and that of the on-body testing is significant.

## Introduction

Wireless technology has been the enabling domain in revolutionizing healthcare system in conjunction with information technology (IT). Technologies such as m-health, ubiquitous monitoring as well as telemedicine have recently become popular and attracted the attention of researchers [1]. Continuous and real-time monitoring are the backbone of many of these technologies which are currently being utilized in healthcare systems [2,3]. Wireless sensor network (WSN), as its name reveals, is an array of sensors (e.g., blood pressure, glucose, ECG) which is networked together wirelessly. These types of networks have found many applications in environmental monitoring, security, sport, tracking and healthcare; the latter is growing rapidly [4] in the past decade. Wireless body sensor networks (WBSNs)- specific-purpose WSNs used in healthcare, have received considerable attention and now support a wide range of applications, among which fall [1,5] electrocardiography (ECG) (i.e., wireless ECG), respiration [6–8] as well as temperature monitoring [5], [9] are listed. Normal human body temperature is approximately 37°C at which all the human organs can effectively function. Unbounded increase of body temperature, particularly in neonates, can cause severe health concerns, and it can adversely affect their normal life sometimes permanently [9]. Long medication process will also be needed if the health concern is severe. Contrary to adults, neonates are subject to increased levels of high body temperature (i.e., fever) side effects.

Usually in hospitals, patients' vital signs (physiological signs) are recorded and supervised by clinical staff during several referrals. Human error, lack of adequate skills, tiredness and inefficient staff in additional to lack of sufficient accuracy (due to wrong interpretation) can deteriorate health of many neonates, especially when the number of hospitalized neonates exceeds. Therefore, considering the aforementioned reasons, the design and implementation of a real-time temperature monitoring system (Figure 1) is found to be essential particularly in ICUs where neonates are placed in case of emergency [10]. In this paper, the design and implementation of a real-¬time monitoring system for the human (e.g., neonates) body temperature is addressed [11]. The system comprises of a number of wireless thermometer and a communication infrastructure for handling data transmission and reception. In comparison to ex¬isting temperature monitoring systems [11–13], the proposed system consists of a built-in identification mechanism. The identification mechanism can distinguish sensor-specific data sent by a wireless thermometer among other measure-ments. Also, the computer software developed in-house for handling and visualizing the measurement results is presented.

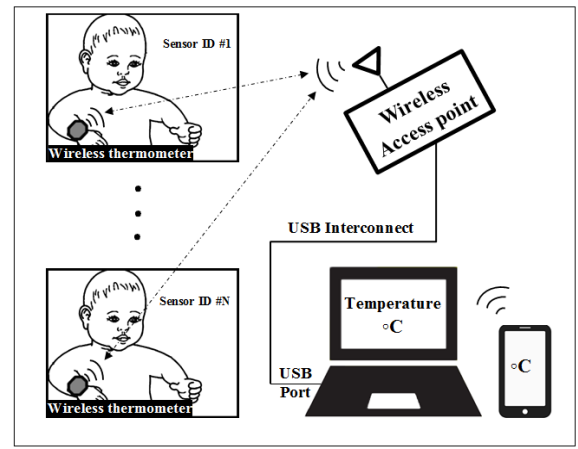

**Figure 1:** Schematic depicting the proposed temperature monitoring system.

### Wireless Thermometer

Figure 2 demonstrates the realized wireless thermometer and the schematic of its internal circuitry. Similar to typical wireless sensor nodes, the proposed thermometer consists of three main parts, namely a transceiver (HM-TR) [14], a microcontroller (Atmega8) [15] and a temperature sensor (DSI8B20) [16]. These parts were fabricated on two separate boards and are connected by using a flat 6-pin interconnect as shown in Figure 2. The microcontroller and the sensor were implemented together on a board. The transceiver in addition to an SMA connector for plugging the antenna (i.e., Monopole antenna) were situated on an individual board, which is com¬mercially available on the market. This configuration made it possible to fabricate the thermometer with reduced size, which is to be worn on neonate's wrist (Figure 1).

### Communication Infrastructure

In order to visualize the temperature values for making necessary alerts and decisions, the thermometers require to wirelessly communicate the readings to a PC or a PDA (Personal Digital Assistant) via an access point which is directly connected to it, as shown in Figure 3(a). The access point composes of a transceiver tuned to operate at 915 MHz, and a R232 to USB port convertor since the transceiver sends out the data (i.e., received temperature value) as sets serial sequence (Figure 3(b)).

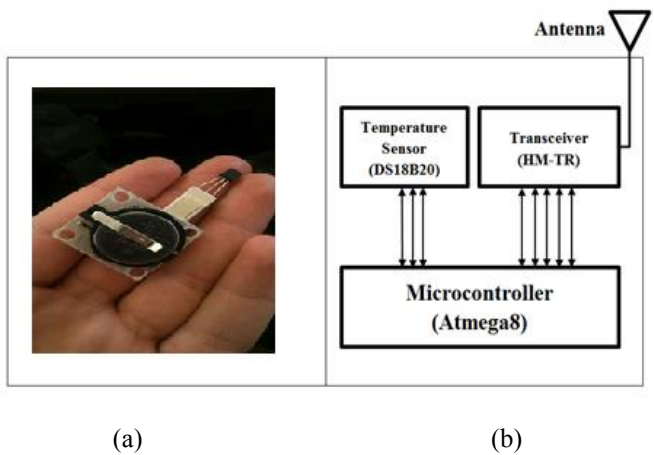

(a)     (b)

**Figure 2:** Wireless thermometer (Wireless temperature sensor node); (a) Picture of the realized thermometer, (b) schematic of the thermometer and the building parts.

### Wireless Medium Access Control (MAC) - TDMA

A number of media access control (MAC) protocols have been proposed for medical sensor networks [5,17–20] Main requirements of a MAC protocol are reliability, flexible transmission mechanism, high channel efficiency and a low end-to-end delay time. Mainly three classes of MAC protocols have been considered for wireless medical applications. They are TDMA (time division multiple access), polling and the contention-based protocols also known as the random access protocols[21–23]. Since more than one thermometer is used in the proposed monitoring system and the sensor nodes need to transmit the data to the access point via a shared wireless medium, a mechanism should be used to control the timing and also to ensure safe access of thermometers to the access point. Otherwise, untethered communication will result in data collision and seemingly loss of information. To do this, a time-based MAC protocol which is called Time division multiple access, was used. The TDMA executes a procedure in which each node first actives a built-in RSSI (Received Signal Strength Indictor) device to estimate the power level in the channel (i.e., shared medium) before deciding to send the data. Once the medium is free, i.e. no communication in progress, the wireless thermometer sends the readings to the access point. TDMA works with principle of dividing time frame in dedicated time slots, each node sends data in rapid succession one after the other in its own time slot. Synchronization is one of the key factors while applying TDMA. It uses full channel width, dividing it into two alternating time slots. TDMA uses less energy than others due to less collision and no idle listening. TDMA protocols are more power- efficient than other multiple access protocols because nodes transmit only in allocated time slots and all other time in inactive state. A packet generated by node suffer three types of delays as it reaches receiver [20]. Figure 4 shows a flow chart for TDMA. In TDMA, first of all each node is assigned a particular time slot for its transmission. Synchronization is done between source node and destination node. Node checks for its particular time slot and transmits data packets in its relevant time slot otherwise it waits for its relevant time slot. If packets are not available for transmission, communication terminates. Otherwise node checks for availability of slot and this process repeats until communication terminates.

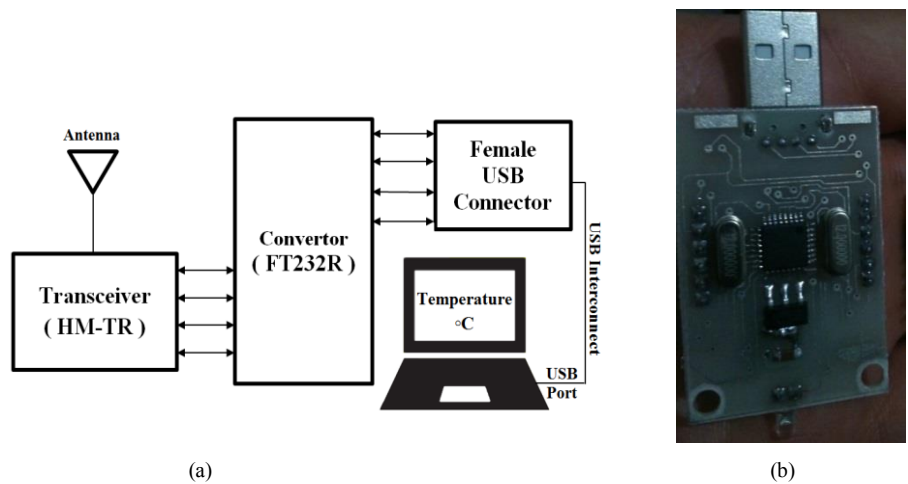

(a)     (b)

**Figure 3:** Communication architecture of the receiving part (access point) (a), the transceiver and an RS232 to USB convertor (b).

### Thermometers Evaluation

The proposed temperature monitoring system comprising of the wireless thermometers and the wireless adaptor (i.e., access point) as well as a personal computer were tested in different scenarios in order to evaluate the performance of the thermometers and also to validate the accuracy of the measurements. In the first experiment, one thermometer was deployed to measure the volunteer's body temperature for 60 sec. with the period of 1 msec. Therefore, the number of measurements recorded by the monitoring system is 60 points. Figure 4 represents the measurement results and as it can be seen, temperature values fluctuate between two upper and lower bands, 30 °C and 26 °C, respectively. For comparison purpose, 2-node wireless thermometer temperature was also measured. The temperature values obtained from the 2 Nodes wireless thermometers are shown in Figure 5. In order to evaluate the MAC protocol and the identification mechanism utilized for distinguishing the readings by the thermometers, two wireless thermometers were set to measure temperature values of two separate objects (Figure 5(a)). The readings of the sensors were collected and visualized by a computer software as shown in Figure 5(b).

### System Delay

In this section, the node delay implementation of the state discussed are examined. Delay of each section is listed in Table 1. The total delay is the total time wasted as obtained, which is shown in Equation (1).

The total delay is the total time wasted as shown in equation (1) is obtained. T(1) delay is the processing of data in the microcontroller. Due to the fact that the microcontroller used in this project is Atmga8, each clock latency from is given by equation (2). The value off is the operating frequency of the microcontroller.

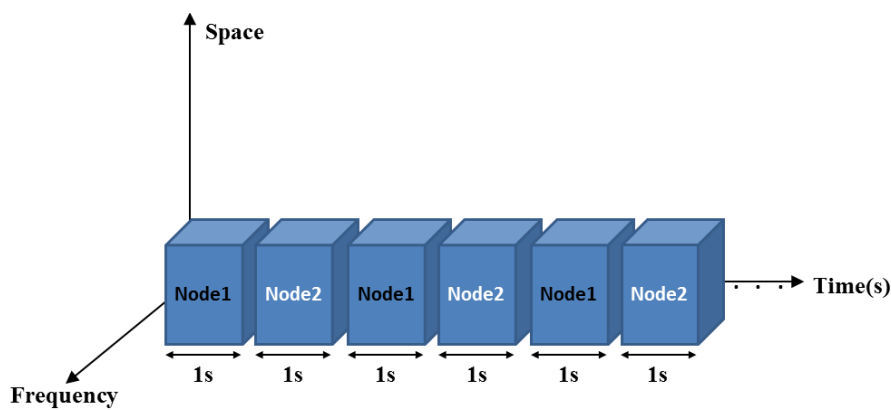

**Figure 4:** Timing diagram of TDMA

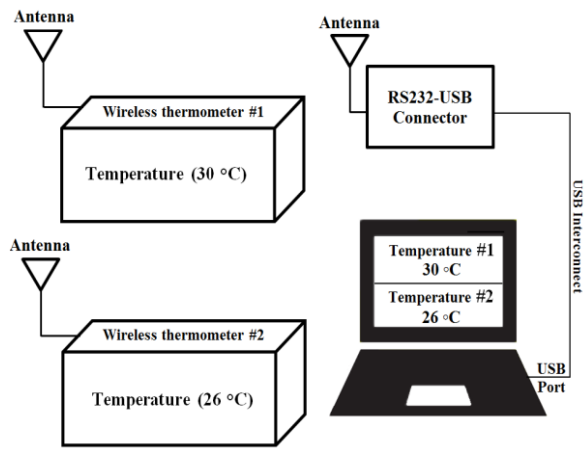
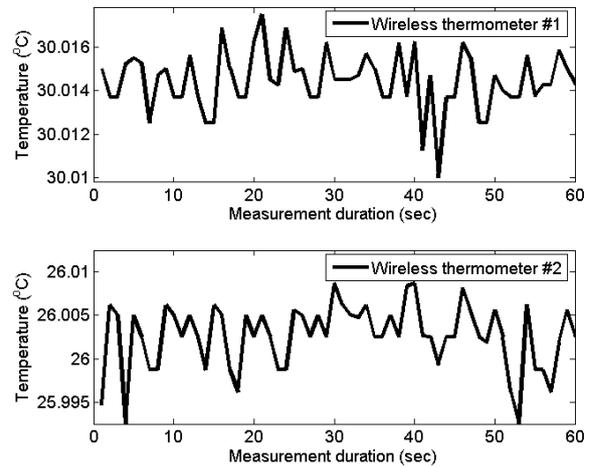

(a)          (b)

**Figure 5:** (a). Schematic illustrating the setup used for testing the array of two wireless thermometers. (b). The temperature values of two separate object measured by the realized wireless thermometer.

**Table 1:** Delay of each section.

| System Delay | |
|---|---|
| T1 | Data preparation microcontroller delay |
| T2 | Delay module from the transmitter to the receiver |
| T3 | The air delay |
| T4 | Delays in receiving data by the central node module |
| T5 | Delay module from the transmitter to the receiver |
| T6 | R232 to USB delay |
| T7 | The temperature sensor delay |
| T8 | USB Wire delay |

$$T(Delay) = T1 + T2 + T3 + T4 + T5 + T6 + T7 + T8 \quad (1)$$

$$T(\text{microcontroller Clock}) = \frac{1}{8000} \quad (2)$$

Microcontroller is needed to calculate the number of clock delays. For calculated delay adding the number of clock to run any command given microcontroller Datasheet, latency is calculated. According to the source, the number of assembly instructions written for TDMA protocols for each sensor is equal to 316 clocks. Microcontroller delay is shown in equation (3).

$$T(1) = \frac{1}{8000} \times 316 \quad (3)$$

Switching delay from the sender to the receiver module, according to Datasheet is 130 microseconds. Equation (4) shows delayed arrival transmitted bits between the sender and the receiver.

$$T(4) = \frac{\text{The number of bits transmitted}}{\text{Air data rate module}} \quad (4)$$

Up data rate of air HMTR module is 19200 bits per second. 256 bits is the number of bits sent. Bit delayed in reaching the air, the transmitting node to the central node receiving data as shown in Equation (5).

$$T(5) = \frac{256}{19200} = 0.013 \text{ms} \quad (5)$$

To calculate the delay R232 to USB, the number of bits transmitted is divided by the FT232 data. By summing up all of the time delay, the total delay a packet to reach its des-tination is characterized. The amount of time needed to convert the digital temperature value is 750 milliseconds. Figure 6 shows delay by increasing the distance and the transmission of bits.

Figure 7 represents the delay by increasing the number of bits as a repeat count. Experiments done nodes within the central node receiving data rate of 10 meter is considered. As shown in Figure 7, the Payload Module transceiver for sending 256 bits, the different amount of delay in the number of Payload bit different.

### Node Energy Consumption

To calculate the energy consumption node, the amount of energy in HMTR module to send, receive and idle power consumption values depend on the microcontroller and sensor DS18B20. Energy can be calculated from Equation (6).

$$E = P \times t \qquad (6)$$

$$P = V \times i \qquad (7)$$

In Equation (7), P as the primary power battery, i amounts consumption during transmission mode and t represents the time to send

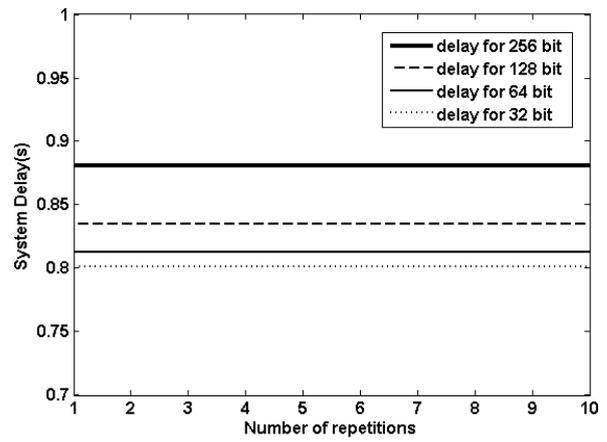

**Figure 7:** Delay by increasing the number of bits.

specified number of bits in packet. Table 2 showing current consumption values of the nodes sender.

Figure 8 shows Energy consumption in state transmitted, Received, Idle with HMTR transmitter module. As seen in Figure 8 increasing bits with repetitions, energy consumption is increasing.

## Conclusion

The design and implementation of a temperature monitoring system were presented. The monitoring system consists of a number of ID-based thermometers, each comprising of a wireless transceiver unit operating at 915

**Table 2:** Current consumption values of the nodes.

| Device | Parameters | Unit | Value |
|---|---|---|---|
| Battery | Primary Power Battery | Watt | 500 |
| Battery | primary voltage battery | Volt | 9 |
| Battery | Initial current battery | Amp | 0.016 |
| HMTR | i(transmit) | Amp | 0.016 |
| HMTR | i(receive) | Amp | 0.036 |
| HMTR | i(Idle) | Amp | 0.000001 |
| DS18B20 | i(Active) | Amp | 0.009 |
| DS18B20 | i(Idle) | Amp | 0.000000008 |
| Atmega8 | i(Active) | Amp | 0.0036 |
| Atmega8 | i(Idle) | Amp | 0.001 |

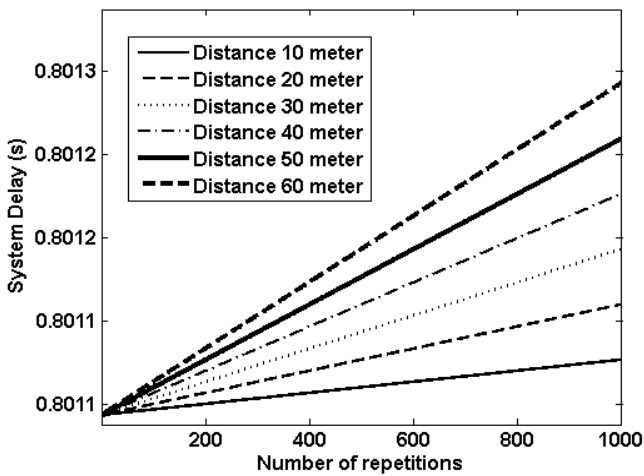

**Figure 6:** Delay by increase the distance and the transmission of bits.

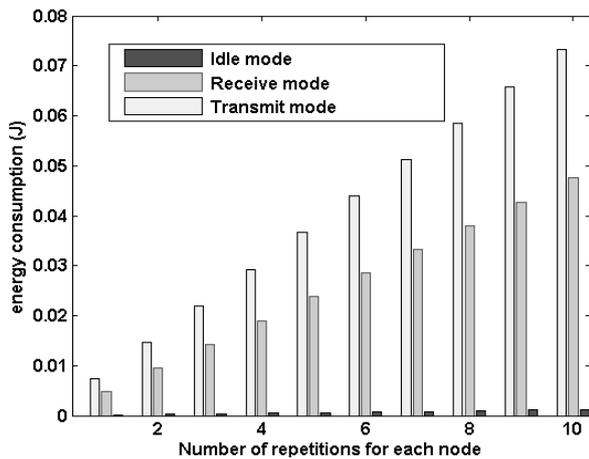

**Figure 8:** Energy consumption (transmitted, Received, Idle).

MHz as well as a temperature sensor which is controlled by a microcontroller.

## Acknowledgment

This research has been supported by the Health Research Center, Baqiyatallah University of Medical Sciences, Tehran, Iran. There is no actual or potential conflict of interest regarding this article.

## Conflict of Interest
None